\documentclass{eas}
\usepackage{graphicx}
\usepackage{epsfig}
\def\gsim{\lower 2pt \hbox{$\, \buildrel {\scriptstyle >}\over
{\scriptstyle \sim}\,$}}
\def\lsim{\lower 2pt \hbox{$\, \buildrel {\scriptstyle <}\over
{\scriptstyle \sim}\,$}}

\def\fuse{{\sl FUSE}}

\def\xmm{{\sl XMM-Newton}}

\def\suzaku{{\sl Suzaku}}
\def\chandra{{\sl Chandra}}
\def\Chandra{{\sl Chandra}}

\def\oi{O~{\scriptsize I}}
\def\oii{O~{\scriptsize II}}
\def\oiii{O~{\scriptsize III}}

\def\ovi{O~{\scriptsize VI}}
\def\ovii{O~{\scriptsize VII}}
\def\oviii{O~{\scriptsize VIII}}

\def\neix{Ne~{\scriptsize IX}}

\def\fexvii{Fe~{\scriptsize XVII}}

\TitreGlobal{Hot Gaseous Halos of Nearby Disk Galaxies }
\begin{document}

\title{Hot Gaseous Halos of Nearby Disk Galaxies} 
\author{ Q. Daniel. Wang}\address{Department of Astronomy, University of Massachusetts, 
  Amherst, MA 01003; wqd@astro.umass.edu}
\begin{abstract}
I review studies of the hot gaseous medium in and around nearby
normal disk galaxies, including the Milky Way. This medium represents a reservoir of 
materials required for lasting star formation, a depository of galactic
feedback (e.g., stellar mass loss and supernovae), and an 
interface between the interstellar and intergalactic 
media. Important progress has been made recently with the detection of 
X-ray absorption lines in the spectra of X-ray binaries and AGNs.
The X-ray absorption line spectroscopy, together with existing X-ray
emission and far-UV \ovi\ absorption measurements now allows
for the first time to characterize the global spatial, thermal, and
chemical properties of hot gas in the Galaxy. The results are
generally consistent with those inferred from X-ray imaging of 
nearby edge-on galaxies similar to the Milky Way.
Observed diffuse X-ray emitting/absorbing gas does
not extend significantly more than $\sim 10$ kpc away from galactic 
disks/bulges, except in nuclear starburst or very massive galaxies. 
The X-ray cooling rate of this gas is generally far less 
than the expected supernova mechanical energy input alone. So 
the bulk of the energy is ``missing''. On the other hand, evidence for a
large-scale ($\gsim 10^2$ kpc) 
hot gaseous halo around the Milky Way to explain various
high-velocity clouds is mounting. The theoretical argument for ongoing 
accretion of intergalactic gas onto disk galaxies is also compelling. I
discuss possible solutions that reconcile these facts.
In particular, large-scale hot gaseous halos appear to be low in
metallicity, hence X-ray emission. The metal enrichment in the
intergalactic medium may be substantially non-uniform; fast-cooling clumps of
relatively high metallicity may have largely dropped out and may partly 
account for high-velocity clouds. In addition, ongoing 
galactic mechanical energy feedback
is likely important in balancing the cooling of the halos and may be
strong enough to produce galactic winds in bulge-dominated galaxies.
\end{abstract}
\maketitle
\section{Introduction}
It is widely believed that disk galaxies were formed and are still evolving 
from the mist of large-scale diffuse gas. Around reasonably 
massive galaxies like the Milky Way, such gas can be heated to X-ray-emitting 
temperatures at the so-called virial shocks (at galactocentric distances of
$\sim 200-300$ kpc) and through subsequent
gravitational compression (e.g., Birnboim \& Dekel 2003). 
This gas, cooling radiatively, can maintain lasting 
star formation in galactic disks. 
But there are serious problems with this pure accretion scenario. Primary 
among these is the so-called ``over-cooling'' problem
(e.g., Navarro \& Steinmetz 1997). This refers to the 
fact that structure formation simulations consistently
over-predict the cooled gas content in galactic disks. Essentially the 
accreted gas,
with a ``typical'' metallicity, would cool too quickly in the forming 
galactic halos. Observationally, the mass of all stars and the interstellar 
medium (ISM) in a galactic disk typically accounts for only about 
half of the baryon matter expected from the accretion. The remaining 
matter, if having not been blown away, presumably resides in the galactic 
halo, most likely in a hot phase (e.g., Fukugita \& Peebles 2006). 
In reality, the accretion is not only sensitive to the galaxy mass, but 
is also  affected by various 
feedback processes such as the pre-heating and metal enrichment 
of the IGM as well as the ongoing energy release from stars 
and/or active galactic nuclei (AGNs; e.g., Toft \etal\ 2002; Mo \&  Mao 2002;
Dav\'e \& Oppenheimer 2006). However, it is not clear which of these processes
is more important. Energetically, the ongoing galactic 
feedback typically provide enough energy to balance the cooling. Indeed,
much of the expected mechanical energy feedback from supernovae (SNe) 
is ``missing'', 
at least not observed in X-ray emission (Wang 2005, Li \etal\ 2006a,b), 
and could meet the energy need for heating the halo. But the 
problem is how to transport the energy from inside a galactic disk/bulge
into a large-scale gaseous halo. 
All these problems are related to the behavior of the 
global diffuse hot gas in and around the galaxies. 
So let me first review the status of characterizing 
the spatial, thermal, and chemical properties of the hot gas
and then discuss possible solutions to the problems.

\section{Hot Gas in and around the Milky Way}

\subsection{Diagnostics}

Half a century ago, Spitzer (1956) postulated the presence of 
a Galactic corona to provide the confinement for H{\scriptsize~I} clouds
observed at high Galactic latitudes. This corona, different from
the large-scale hot gaseous halo as expected from the IGM accretion,
was thought to be heated by the Milky Way, similar to the relationship
between the Sun and its corona.
Direct evidence for the Galactic corona  at a temperature of $\sim 10^6$~K,
of course, needs to come from X-ray and far-UV observations. 
Various broad-band X-ray 
observations have indeed revealed an extensive diffuse soft  ($\lsim 1$~keV)
X-ray background 
(SXB; e.g., Fig.~1, Snowden \etal\ 1997; 2000). High spectral resolution X-ray 
emission data, made with sounding rockets and accumulated from 
large swaths of the sky (McCammon \etal\ 2002), have further shown
emission lines such as C{\scriptsize~VI}, O{\scriptsize~VII}, and O{\scriptsize~VIII}, confirming 
the thermal origin for much of the SXB.
X-ray emission observations alone, however, give little distance information. 
One may infer the location of X-ray-emitting plasma relative
to X-ray-absorbing cool gas, by comparing their relative distributions
in the sky, though often with great uncertainties. A substantial part of 
the SXB, even in the 1/4-keV band, appears to arise from regions 
beyond the immediate solar neighborhood, or the so-called Local Bubble (LB). 
In particular, there is a general intensity enhancement towards the inner 
region of the Galaxy in the 3/4-keV and 1.5-keV bands. This enhancement
is at least partly due to the emission from the Galactic bulge and
possibly from a hot gas outflow from the Galactic center. 
The intensity distribution is substantially patchier in the 1/4-keV 
band than in the 3/4-keV band, which cannot be entirely due to absorption 
by cool gas (e.g., Kuntz \& Snowden 2000). There is also no 
intensity correlation between the two bands. These differences strongly
suggest the presence of gas components at quite different temperatures 
(e.g., $\sim 10^6$ K vs. $\sim 3 \times 10^6$ K); the hotter gas with a 
longer cooling time scale tends to be more abundant and diffuse than 
the cooler one.
But, it is still not clear whether the Milky Way actually has a smoothly
distributed Galactic corona, which may explain the bulk of the 3/4-keV
diffuse emission (Wang 1997; Pietz et al. 1998), or just a composite of various discrete
hot gas outflows or Galactic fountains from the Galactic disk as well as 
the bulge.

\begin{figure}[h]
\includegraphics[width=0.5\textwidth,angle=90.0]{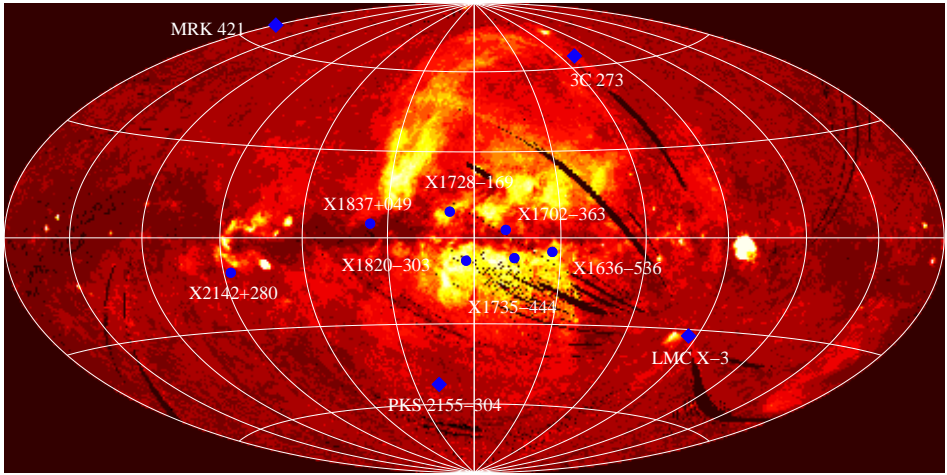}
  \caption{\scriptsize {\sl ROSAT} all-sky survey map of the diffuse 3/4-keV-band 
background intensity (in the Aitoff projection; Snowden \etal\ 1997). 
The blue symbols mark the directions of X-ray sources 
toward which X-ray absorption lines
by Galactic hot gas have been detected (e.g., Yao \& Wang 2005; 
Wang \etal\ 2005).
 }
\end{figure}

Hot gas in the Galaxy is also traced by far-UV absorption/emission
lines, mostly observed in 
absorption. The vertical scale heights of \ovi, N{\scriptsize~V}, C{\scriptsize~IV}, 
and Si{\scriptsize~IV} absorptions,
as inferred from their distributions across the sky, are 
{\mbox{$\sim$2.3,}} 3.9, 4.4, and $5.1$ kpc, respectively (e.g., Savage \etal\ 2000; 
Savage \etal\ 2003), clearly showing a trend toward lower ionization 
states at larger distances from the Galactic plane. This trend 
 may be understood as a result of Galactic fountains,
 in which hot gas heated primarily by SN blastwaves in the
 Galactic disk cools off on the way to the Galactic halo.
While the populations of C{\scriptsize~IV} and Si{\scriptsize~IV}
could be largely due to photon-ionization, \ovi, 
sensitively probed with {\sl FUSE} as the
1031.9  and 1037.6 \AA\ resonance line doublet, should originate 
predominantly in
collisionally ionized plasma. Indeed, the doublet has also been 
detected in diffuse emission, which requires very long exposures with
\fuse\ because of its small aperture ($30^{\prime\prime} \times
30^{\prime\prime}$). Unlike the absorption, however, the 
emission intensity is subject to extinction, the correction 
of which can be rather uncertain. Existing
emission observations are also typically not taken in close vicinities of 
the absorption sight-lines. Therefore, the comparison between 
the absorption and emission data has been difficult. Furthermore, the 
inference of the relevant physical parameters from the far-UV data alone 
requires an assumption
of the gas temperature, often taken to be $\sim 3 \times 10^5$~K, at which
the  \ovi\ ionic fraction peaks in collisionally ionized gas. Under this 
assumption, one neglects the \ovi\ contribution from thermally more stable and
thus more abundant gas at higher temperatures.

The most direct and powerful tool to study the {\sl global} properties of
hot gas in the Galaxy is the X-ray absorption line spectroscopy
(e.g., Yao \& Wang 2005, 2006a,b). The X-ray
bandpass contains all the K-shell transitions of  
carbon through iron and the L-shell transitions of silicon through iron
(Paerels \& Kahn 2003).
The grating instruments aboard {\sl Chandra}  
and \xmm\ X-ray Observatories, with a spectral resolution of 
$\sim 500 {\rm~km~s^{-1}}$, now enable us to study the hot gas by detecting its
line absorption in the spectra of bright background X-ray sources. 
The strongest lines are produced by ions such as \ovii, \oviii, and \neix,
which sample the entire expected temperature range of the hot gas 
($\sim 10^{5.5}- 10^{6.5}$ K) in
collisionally ionization equilibrium (CIE; Yao \& Wang 2005).
Unlike X-ray emission, which is sensitive 
to the volume density, the absorption
lines trace the column density, 
which is proportional to the mass of the hot gas. Detection of 
multiple lines along the same sight-line can further provide sensitive
diagnostics of the thermal, chemical, and kinematic properties of the hot gas. 
These diagnostics are independent of ionization-edge 
(photo-electric) absorption dominated by abundant interstellar cool 
gas ($T \lsim 10^4$ K) and are therefore ideal for studying the global 
hot ISM.
Comparison of the absorption lines detected along two or more sight-lines
further allows to examine the differential properties of the hot ISM.
Combining the absorption and emission measurements, we can 
even estimate the effective depth of hot gas along individual sight-lines.
Existing results from using these capabilities are reviewed in the following.

\subsection{Recent Results}

\medskip
\noindent{$\bullet$} Spatial Extent

The X-ray absorption lines produced by local hot gas ($cz \sim 0$) are first detected 
in the spectra of several bright AGNs (e.g., Fig.~2). Typically 
only the \ovii\ K$\alpha$ line is
detected, often marginally, except for PKS~2155--304, 3C~273, and 
Mrk~421 (e.g., Rasmussen \etal\ 2003; 
McKernan \etal\ 2004; Williams \etal\ 2005; Fang \etal\ 2006 and 
references therein). Early analyzes all assume a single temperature
for the absorbing gas, which is certainly an over-simplification. 
Williams \etal\ (2005) show that the one-temperature assumption is only 
consistent with an extragalactic origin of the gas along the sight-line
toward Mrk 421 (e.g., in the Local Group).
However, the detection of the \ovii\ K$\alpha$ line with a similar
equivalent width in the spectrum
of LMC X-3 at a distance of 50 kpc strongly suggests that the absorption 
is primarily Galactic in origin (Wang \etal\ 2005).
Stronger absorptions are also detected in the spectra of Galactic 
low-mass X-ray binaries (LMXBs), when the signal-to-noise ratios of the \chandra\ 
observations are reasonable high (Futamoto \etal\ 2004; 
Yao \& Wang 2005).
Most of the observations were taken for the study of the objects themselves
(therefore the use of the high-energy transmission grating; HETG) 
and were not necessarily optimized for the ISM study. The HETG spectra most commonly show the \neix\ K$\alpha$ absorption line. A joint analysis of
the detected absorption lines along the Galactic and extragalactic sight-lines
further suggests that the absorbing gas is located around the Milky Way
and with effective scales no more than a few kpc, depending on the 
assumed overall morphological shape, disk- or sphere-like (Yao \& Wang 2005;
Juett \etal\ 2006).
This, of course, does not exclude the presence of hot gas in a substantially
larger region around the Galaxy (see \S 4.2 for more discussion). 
But it is clear that the bulk of the observed
X-ray emission and absorption arises in the immediate vicinity of 
the Galactic disk/bulge.

\medskip
\noindent{$\bullet$} Temperature and Density Distributions
\begin{figure}
\includegraphics[width=1\textwidth,angle=0.0]{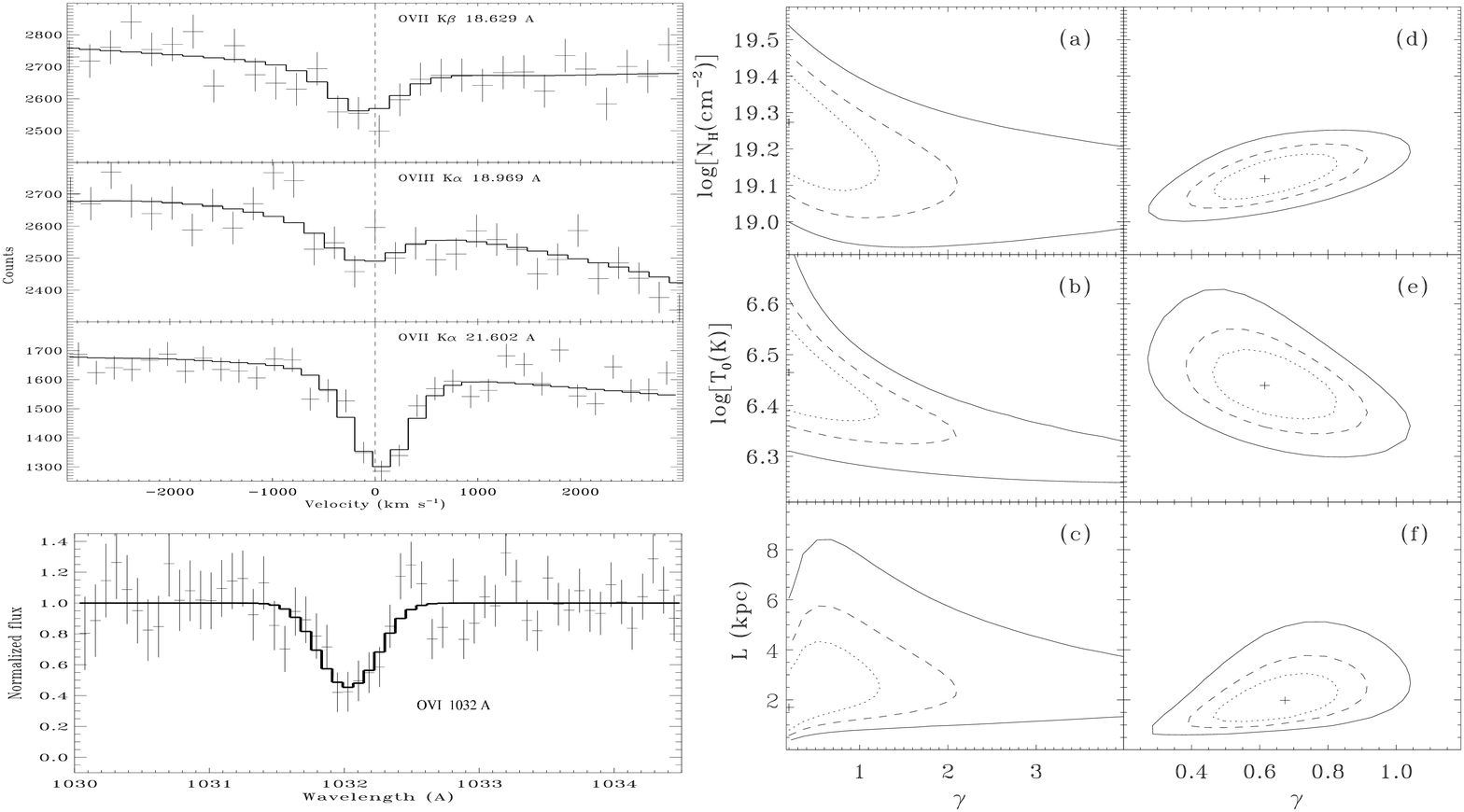}
   \caption{\scriptsize Left panels: \chandra\ detection of \ovii\ K$\alpha$ and
K$\beta$ and \oviii\ K$\alpha$ absorption lines and the \fuse\ detection 
of the \ovi\ 1031.9
\AA\ absorption line along the Mrk 421 sight-line.
The histograms represent the fitted model, assuming a
power-law hot gas temperature distribution  (d$N_{\rm H}/{\rm d}T \propto 
T^{\gamma-1}$; Yao \& Wang 2006b). Right panels:  
The 68\%, 90\%, and 99\% confidence contours of the model parameters: 
   the total hot hydrogen column density
 $N_{\rm H}$, the maximum (mid-plane) temperature $T_{\rm 0}$, 
 and the effective path-length $L$ 
  versus $\gamma$. The constraints are obtained from the joint fits
 to the X-ray absorption data and the \ovii\ and \oviii\ emission 
line measurements from McCammon \etal\ (2002) with 
    (d-f) or without (a-c) the inclusion of the 
    \ovi\ absorption line (Yao \& Wang 2006b). 
   \label{fig:mrk421} }
\end{figure}

Along the sight-line to Mrk~421, Yao \& Wang (2006b) find that the measured 
absorption line strengths of \ovii\ and \oviii\ (Fig.~2) are 
{\sl inconsistent} with the diffuse emission line ratio of the same ions, if 
the hot gas is assumed to be isothermal in a CIE
state. But all these lines as well as the diffuse 3/4-keV broad-band background
intensity in the field can be jointly fitted with a plasma with a
power-law temperature distribution, which 
can be derived from a vertical exponential disk model 
of hot gas. The joint fit gives the 
exponential scale heights of the density and temperature as $\sim 1.0/f$ kpc 
and $1.6/f$ kpc and their mid-plane values as $2.8\times10^6$ K 
and $2.4 \times 10^{-3}\ {\rm cm^{-3}}$, respectively. The filling factor 
of the hot gas $f$ is still very uncertain, although a preliminary 
multi-phase column density comparison for the 4U~1820--303 sight-line
indicates that the filling factor of hot gas is $\gsim 0.8$.

\medskip
\noindent{$\bullet$} Metal Abundances 

So far the most comprehensive study of interstellar X-ray absorption lines
is for the sight-line toward the Galactic bulge LMXB 4U~1820--303, located in 
a globular cluster
(Galactic coordinates $l, b=2^\circ.79, -7^\circ.91$ and distance = 7.6 kpc). 
The {\sl Chandra} grating spectra of the source show absorption lines produced by
\oi, \oii, \oiii, \ovii, \oviii, \neix, and \fexvii. The detection of 
these lines allows us to measure 
the column densities of the cold, warm, and hot atomic phases of the ISM 
through much of the Galactic disk (Yao \& Wang 2006b, Yao \etal\ 2006). 
The hot phase of the ISM 
accounts for about 6\% of the total atomic oxygen column density
$8.0 \times 10^{17} {\rm~cm^{-2}}$ along the sight-line.
By comparing these measurements 
with the 21 cm hydrogen emission and with the pulsar dispersion measure 
along the same sight-line, we have further estimated the mean oxygen abundances in the 
neutral and ionized (warm plus hot) phases as 0.3(0.2, 0.6) and 2.2(1.1, 3.5) solar 
(90\% confidence intervals).
We have also obtained the Ne/O and Fe/Ne abundance ratios of the hot phase as 1.4(0.9, 2.1) 
and 0.9(0.4, 2.0) solar. The results indicate a substantially increased molecule/dust grain 
destruction and/or enhanced metal enrichment in the warm and hot phases of the ISM. 

\medskip
\noindent{$\bullet$} Relationship between \ovi- and \ovii-bearing gases

A hint about this relationship comes from the similar kinematic properties of
\ovi- and \ovii-bearing gases. Although
the X-ray absorption lines are not resolved in the existing 
grating spectra, we can estimate the velocity dispersion of the absorbing
gas from the \ovii\ K$\alpha$/K$\beta$ ratio. The estimated average
dispersion is $\sim 85(74, 116) {\rm~km~s^{-1}}$ for the LMC X-3 and Mrk 421
sight-lines, which sample the local hot gaseous disk/halo at 
high Galactic latitudes (Wang \etal\ 2005; Yao \& Wang 2006b), 
and $\sim 255(165, 369) {\rm~km~s^{-1}}$ for 4U~1820--303 toward 
the Galactic bulge (Yao \etal\ 2006). The 
high velocity dispersion of the hot gas along this latter sight-line
is probably due to the large differential rotation, which would occur
mostly in the Galactic Center region (Clemens 1985), 
and/or to the enhanced bulk/turbulent 
motion expected in the Galactic bulge region. Even
the dispersion in the local disk/halo is significantly greater than
the value expected from the thermal motion of the \ovii\ ions 
 ($\sim 36 {\rm~km~s^{-1}}$) and is 
consistent with those ($\sim 80-90 {\rm~km~s^{-1}}$; e.g., Wang \etal\ 2005) 
directly resolved in the \fuse\ \ovi\ absorption line observations
with a resolution of $\sim 20 {\rm~km~s^{-1}}$.  
The \ovi\ and X-ray absorption lines
further show little offset from the local standard of rest.
These similarities and consistencies indicates that the \ovi\ absorbers 
are dynamically mixed with the X-ray-absorbing hot gas and originate
mostly in a thick rotating hot disk with a weak bulk/turbulent 
motion, probably due to a Galactic fountain (Bregman 1980; Wang \etal\ 2005).
So it is reasonable to assume that at least part of the \ovi-bearing
gas arises from the cooling X-ray-absorbing hot gas and hence may be 
characterized with a single temperature distribution.
The inclusion of the {\sl FUSE} \ovi\ absorption data can greatly tighten
the constraints on the temperature distribution of the hot gas (Yao \& Wang 
2006b). 
Fig.~2 (a-f) compares the confidence contours 
of the parameter constraints with or without the 
inclusion of the \ovi\ absorption line in the fit for the Mrk 421 sight-line. 

We may further explore the relationship between the \ovi- and \ovii-bearing gases by
comparing the absorption lines with the measurements
of the background \ovi\ line emission. A preliminary comparison shows
that the temperature distribution inferred from the absorption line fits 
for the local disk/halo sight-lines predicts an \ovi\ emission intensity 
that is substantially smaller (by a factor of $\sim 4$)
than  $\sim 4000  
{\rm~photons~s^{-1}~cm^{-2}~str^{-1}}$, typically measured 
with {\sl FUSE} observations of high S/N ratios and at high Galactic latitudes
(e.g., Shelton 2002; Dixon \etal\ 2006). Therefore, the bulk of the \ovi\ emission
likely arises from an additional component with a characteristic 
temperature of $\sim 3 \times 10^5$~K. 
This component, most naturally representing the combined contribution from
conductive interfaces around cool gas clouds (e.g., Borkowski \etal\ 1990), should have 
a relatively high mean density or a small volume filling factor, compared
to the diffuse hot gas in which they are embedded. The presence of
this discrete component of \ovi-bearing gas also helps to explain its
large column density variation from one sight-line to another. 
But more studies of
existing and upcoming far-UV/X-ray emission and absorption observations
are needed to quantify the component. Now assuming that only 
part of the observed \ovi\ absorption is associated with the diffuse 
hot gaseous disk, the contours shift primarily to the right  in Fig.~2 (d-f). 

Clearly, the existing characterization of the global hot ISM is still
very limited. Much more can be learned even with the existing
capabilities of \chandra, \xmm, \suzaku, and \fuse. Detailed
studies of individual sight-lines and their differential properties will 
ultimately
enable us to characterize the dependence of the thermal, chemical, and
kinematic properties of the global hot ISM on the Galactic radius
as well as the vertical off-plane distance. Of course,
our inside view of the hot gas in and around the Milky Way
is further complemented by various external
perspectives we now have on nearby disk galaxies of various types.

\section{Hot Gas around Nearby Disk Galaxies}

Thanks largely to \chandra\ and \xmm\ X-ray observatories, much progress has also been made recently 
in the study of hot gas around nearby normal spiral galaxies, 
especially edge-on ones (Wang \etal\ 2001, 2003;
Wang 2005, Li \etal\ 2006a,b). With superb spatial resolution, 
{\it Chandra} observations in particular 
allow to cleanly detect extraplanar diffuse hot gas 
and to explore its relationship to other galactic components. 
About 10 such edge-on galaxies have been observed with relatively
deep exposures ($\gsim 50$ ks), although only a few of them have
been adequately analyzed. Some preliminary characterizations of 
the diffuse X-ray emission from the galaxies are as follows:

\medskip
\noindent{$\bullet$} Morphological Properties

The diffuse X-ray emission appears to consist of two main components:
one is correlated with the star formation (SF) rate and the other with the
mass of a galaxy. The SF component tends to have an elongated  morphology along
the galactic disk
(Fig.~3). Both the overall morphological shape and extent are 
similar to those observed in $H\alpha$ and in radio continuum. 
The mass-related component is readily seen in early-type disk galaxies (Type 
Sb-Sa) with little SF. Examples are NGC~4565 (Wang 2005) 
and M104 (Sombrero; Fig.~4a). The
morphology of the X-ray emission resembles the stellar light distribution, 
typically concentrated toward galactic bulges. But the X-ray emission 
can be far more extended than the stellar light (e.g., Fig.~4b). 
Also little correlation is found between X-ray and radio emission
of such galaxies. The X-ray-emitting gas is most likely heated
by Type Ia SNe. A quantification
of the exact spatial scale of the relatively faint diffuse X-ray emission 
around normal disk galaxies is 
difficult and has hardly been done consistently, due to uncertainties in
the component separation, discrete source subtraction, 
background removal, and 2D-to-3D de-projection. Nevertheless, the effective 
(exponential) scales of the emission all seem to be in a range 
of 1-4 kpc (e.g., Li \etal\ 2006, Strickland \etal\ 2004).

The extraplanar hot gas, responsible for the diffuse X-ray emission, 
appears to have substantial substructures, which are best
appreciated in disk galaxies that are moderately inclined. Fig.~3b shows
that the diffuse X-ray emission arises 
primarily from the inner galactic disk of the Sb galaxy NGC~2841,
and extends chiefly toward the southwest. 
This morphology is apparently a result of
various hot gas plumes sticking out from the disk (with its northeast 
portion tilted toward us) and into the halo of the galaxy. These  
X-ray-emitting plumes extend vertically up to a few kpc and
most likely represent blown-out hot gas heated in the galactic bulge and 
massive star forming regions in the disk.

\begin{figure}[bth]
\vspace{-0.7cm} 
\begin{center}
\hbox{\hspace{-0.3cm}\vspace{-1.0cm}
\includegraphics[width=1\textwidth,angle=0.0]{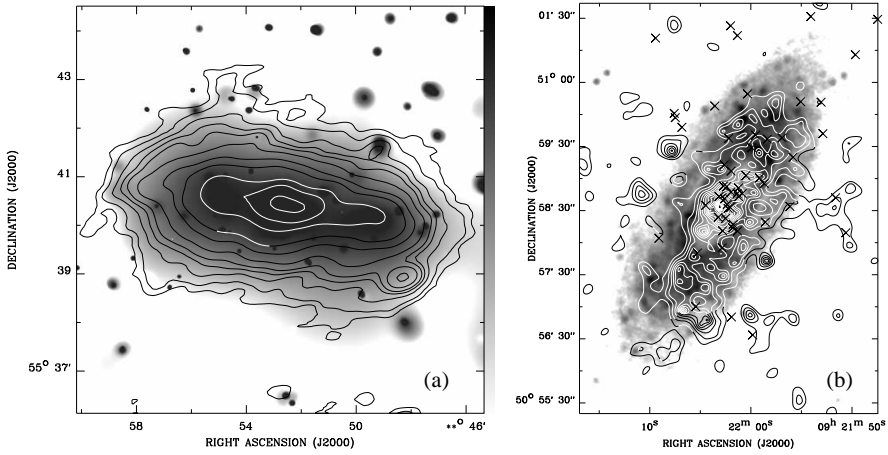}
}
\end{center}
\caption{\scriptsize (a) \chandra\ ACIS-S 0.5--1.5~keV intensity
image of NGC~3556 with overlaid radio continuum 
contours (Wang \etal\ 2003).
(b) ACIS-S 0.5--1.5~keV intensity contours, 
overlaid on an optical blue image  of NGC~2841. 
}
\vspace{-0.5cm}
\end{figure}

\begin{figure}[h] 
\centerline{
\includegraphics[width=1\textwidth,angle=0.0]{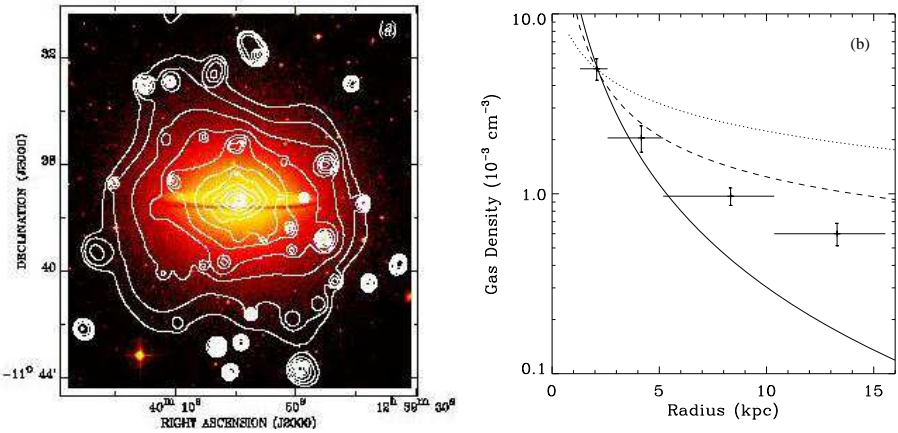}
}
\caption{\scriptsize (a) \xmm\ EPIC-PN 0.5-2 keV intensity contours overlaid on the 
digitized sky-survey blue image of M~104. (b)
De-projected radial density distribution of hot gas {(\sl crosses}),
compared with the model predictions, assuming an adiabatic (dotted curve)  
or isothermal (dashed) gaseous corona in hydrostatic equilibrium or
a 1-D steady galactic wind (solid). 
}
\label{fig:m104}
\end{figure}

\medskip
\noindent{$\bullet$} Spectral Properties

The diffuse X-ray spectrum can be characterized typically with a thermal 
plasma. The characteristic temperature is in the range of 
$(1-7) \times 10^6$ K, depending on both the SF rate and mass of
individual galaxies. The metal abundances of the thermal plasma 
appear to be enhanced:
O-like elements in late-type spirals and Fe-like elements in early-type
spirals. But in general, 
the metal abundances are not
well constrained, subject to uncertainties in the background subtraction
and the assumed temperature distribution of the plasma.

\medskip
\noindent{$\bullet$} Energetics

The X-ray luminosity of the diffuse X-ray emission is proportional to the SF 
rate and to the stellar mass, as traced by the far-IR and K-band luminosities 
of a galaxy. But the luminosity typically 
accounts for only a few percent of the SN
mechanical energy input expected from simple empirical estimates. 
This missing energy problem 
becomes particularly acute in so-called low $L_X/L_B$ 
bulge-dominated galaxies (typically Sa spirals, S0, and low mass ellipticals). 
In a relatively deep \chandra\ observation, 
the bulk of the X-ray emission from such a galaxy is resolved into 
point-like sources (e.g., LMXBs);
the remaining ``diffuse'' X-ray component generally shows a soft spectrum, 
indicating a primarily thermal origin (Irwin \etal\ 2002; O'Sullivan \etal\ 
2003; Wang 2005). But the luminosity of this component
accounts for no more than a few percent of the expected Type Ia SN mechanical
energy input alone. 

In principle, much of the mechanical
energy input could be released by gas at lower temperatures and hence is not
observed as X-ray emission. In a pilot study, Otte \etal\ (2003) 
detected the
\ovi\ 1031.9\,\AA\ emission line in two regions of the NGC\,4631 halo. Follow-up
observations have detected the line in several other regions of the halo (E. Murphy,
private communications). The line centroids of the
detected O{\scriptsize~VI} emission appear to match the underlying disk rotation velocities
reasonably well, indicating an origin in cooling galactic fountains or chimneys 
(Otte \etal\ 2003). The velocity widths of the O{\scriptsize~VI} lines are
considerably greater than the expected thermal broadening, but could
be accounted for by the line-of-sight velocity dispersion
of the galactic disk and/or by the turbulent motion of the gas.
The inferred cooling rate of the \ovi-bearing gas does not seem to be 
greater than the thermal X-ray luminosity of the galaxy.
Interestingly, O{\scriptsize~VI} line emission is not detected from NGC~891 and from  
superwinds around nuclear starburst galaxies such as M82 (Otte \etal\ 2003;
Hoopes \etal\ 2003).
This non-detection may, however, be explained by the relatively 
high foreground extinctions and short {\sl FUSE} exposures of these 
targets. Deeper observations are clearly desirable to
understand the role that \ovi-bearing gas plays in the thermal evolution of
hot gas.

\section{Discussion}

The results summarized above represent a basic characterization of
hot gas in and around nearby normal disk galaxies. Two outstanding 
problems become apparent: 1) The observed far-UV/X-ray emitting/absorbing gas 
close to the galactic disks/bulges is heated primarily by SNe;
but the inferred radiation luminosity of the gas accounts for only a few 
percent of 
the expected mechanical energy input. 2) There is little evidence for the 
large-scale X-ray emission or 'over-cooling'' of the accreted IGM, 
as predicted in galaxy formation simulations. In the following, I 
discuss possible solutions to these two problems.

\subsection{The Missing Stellar Feedback Problem}

One possibility is that the missing stellar energy feedback
is gone with galactic winds. In early-type spirals or elliptical
galaxies with little SF, such winds are likely driven by Type Ia SNe. 
Theoretical studies of galactic-scale gas flows have had a long history 
(Mathews \& Brighenti 2003 and references therein). Various 
solutions, mostly 1-D, have been examined, ranging from
steady to time-dependent ones and from inflow to outflow, depending on
the mass of an individual galaxy. 
But, comparisons of these solutions with observations are so far very much 
limited to such integrated quantities as the galaxy-wide 
X-ray luminosities. There has been
little direct confrontation of the model-predicted 
spectral and spatial properties of diffuse hot gas with 
current X-ray observations.

We have been conducting both 1-D and 3-D studies of Type Ia SN-driven 
galactic bulge winds. We have first constructed
a simple spherically symmetric wind model (Li \& Wang 2007). 
We assume that the mass and energy inputs are proportional to the K-band
intensity distribution and account for the galactic gravitational potential of 
stellar and dark matters in determining the gas dynamics. 
This model is a function of three basic parameters: the stellar mass loss and
energy input rates as well as the galactic bulge size. A preliminary 
comparison of the predicted density profile with a de-projection
measurement based on \chandra\ and \xmm\ data is shown in 
Fig.~4b (Li \etal\ 2006b). The measurement assumes that the gas is
in CIE. The apparent deviation of the measurement from 
the wind model prediction indicates that this assumption may not be valid.
Indeed, one may expect that the fast adiabatic cooling of the wind
results in delayed recombination (Breitschwerdt \& Schmutzler 1999; 
Ji, Wang \& Kwan 2006).
This non-ionization equilibrium (NIE) effect can naturally lead to both
a hardened X-ray spectrum and a relatively flat X-ray surface 
intensity distribution at large radii, compared to the 
CIE predictions. We are currently carrying out quantitative NIE
calculations of the bulge winds. 

\begin{figure}[htbp]
\includegraphics[width=0.5\textwidth,angle=0.0]{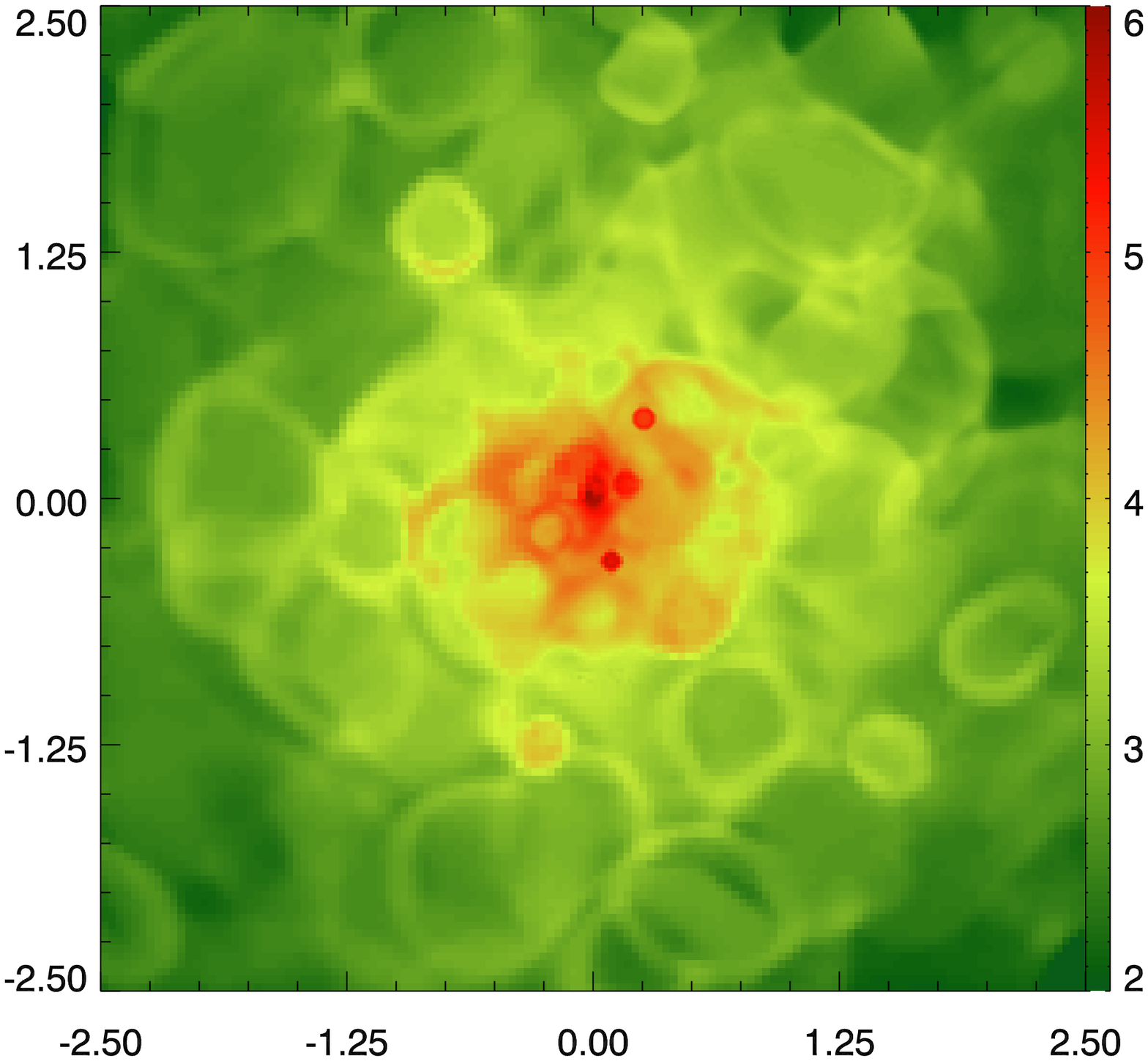}
\includegraphics[width=0.5\textwidth,angle=0.0]{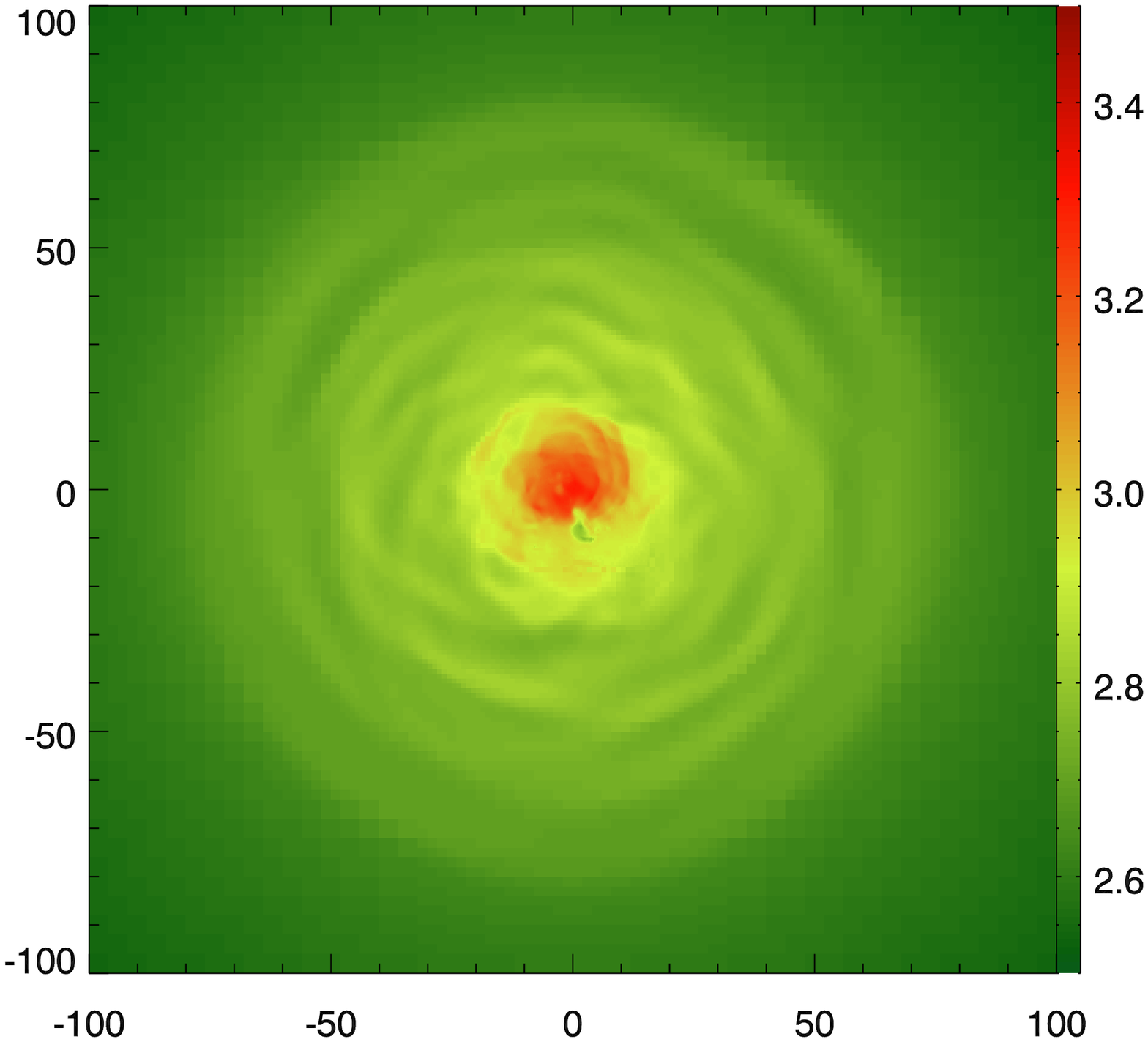}
\caption{\scriptsize \emph{Left panel:} A slice of a 3-D simulated gas thermal pressure 
distribution in a galactic bulge 
similar to the one in the Milky Way or
M31, assuming a wind solution.  This slice across the galactic 
center shows various shell
structures that correspond to enhancements caused by SNe. 
Weak shocks are seen as light green rings. The highest spatial
resolution is 6\,pc.
\emph{Right panel:} A similar pressure slice 
on a large scale from a test run, assuming 
a pre-existing hydrostatic halo.  The ring
patterns show that weak waves excited in the bulge propagate into the halo.
The X and Y axes are all in units of kpc.} 
\end{figure}

We also test the suitability of the 1-D model and possibly calibrate it,
by conducting 3-D simulations for representative galactic bulges. 
These simulations are used to characterize the effects of
key 3-D physical processes and phenomena (e.g., discrete SNR heating
and turbulence), which should not be sensitive 
to the exact global galactic properties assumed.
The characterization of the 3-D density, temperature,
and chemical structures is particularly important for 
comparison with X-ray observations and for determining
the fate of SN chemical enrichment. 
We have already had a basic setup for such simulations.
We use the parallel, adaptive mesh refinement FLASH hydro-code to 
cover a large dynamic range of the structures and to 
effectively capture shocks. 
SNe are randomly generated in time and according to the spatial distribution
of stellar light. Our simulations (e.g., Fig.~5a)
clearly show that the hot gas is full of dynamic structures,
which could not be studied in the 1-D wind model. A preliminary
comparison shows that the corresponding 1-D
solutions substantially underestimate the X-ray fluxes at both 
low and high photon energies, due to the lack of the broad 
temperature distribution shown in the 3-D simulations. 
Furthermore, only massive galactic bulges 
are able to generate winds that are energetic enough to overcome both
the gravity of the galaxies and the in-fall of the IGM. 

For a small galactic bulge such as the one in the Milky Way,
one instead expects convective flows, mixing 
outflows with accreted IGM. Much of the ongoing galactic 
feedback energy can also propagate outward in waves (Fig.~5b;
Tang \& Wang 2005, Tang \etal\ 2007), which can steepen 
into dissipative shocks, especially in and around cooling gas clumps (with 
decreasing sound speeds) formed in the hot gaseous halos. 
Such intra-halo ``tsunamis'' provide a promising mechanism for substantially
slowing down the cooling of the accreted gas. 

\subsection{The Over-cooling Problem}

While no significant observational evidence is found for the ``over-cooling'' 
of hot gas in and around galactic disks/bulges on scales of 
$\lsim 10$ kpc, there is also little sign for diffuse X-ray-emitting 
halos on larger scales (Wang 2005; Li \etal\ 2006b). One probable 
exception is the claim made by Pedersen \etal\ 
(2006) for the detection of an apparent X-ray-emitting halo 
($r \sim 20$~kpc) around the quiescent 
galaxy NGC~5746, based on a moderate 
exposure (37 ks) with the {\sl Chandra} ACIS-I. They interpret this halo as
the evidence for the hot gas from the IGM accretion. However, 
a careful re-analysis of the same 
\Chandra\ data shows no significant detection of such a halo, 
after accounting for the position (mainly radial) dependence of the soft X-ray 
detection sensitivity due to the ice-building on the optical blocking 
filter of the ACIS-I detector, which was not corrected for in the early
version of the analysis software used by Pedersen \etal\
(2006). In addition, part of the claimed halo
may also be due to an inadequate removal of point-like sources.
Of course, even if a truly diffuse X-ray-emitting gaseous halo is 
convincingly detected, it may still be due to the ongoing galactic 
feedback, SF- and/or stellar mass-related, as discussed above.
On the other hand, the presence of a large-scale hot gaseous halo, though
apparently very weak in X-ray emission, is required to confine high-velocity 
clouds seen at large distances from the Galactic disk and to explain their 
apparent ram-pressure stripping morphologies and conductive 
interfaces (e.g., detected in O{\scriptsize~VI} absorption lines; Peek \etal\ 2006
and references therein). 

A scenario that may reconcile these observations with the
galaxy formation theory is that much of the halo gas is very low 
in metallicity.  Metals are known to be present in the IGM; the 
intra-cluster medium with a typical metal abundance of $\sim 1/3$ solar 
is often used as an example. But this view of the 
metal enrichment is strongly biased, and the average metal
abundance of the IGM in the field and in galaxy groups is expected to be 
significantly lower (Dav\'e \& Oppenheimer 2006). Most importantly, 
the enrichment
is probably very non-uniform even on scales substantially smaller than
the sizes of individual galaxies. Unlike in a galactic disk, where shear flows due
to differential rotation can lead to efficient gas mixing, 
the IGM is not expected to be very turbulent.
The mixing or diffusion of metals from early galactic feedback into the IGM
has to be complete on microscopic scales to affect its cooling. 
Therefore, one may expect that the metal enrichment is highly clumpy
in the IGM. Such medium, after being
accreted into the Galactic halo, would cool in a rate 
proportional to the local metallicity. While metal-rich gas quickly 
condense into clouds and fall toward the Galactic disk, 
the remaining medium tends 
to have zero or low metallicity, especially in inner regions of 
the Galactic halo. This scenario qualitatively explains the lack of a
giant X-ray-absorbing/emitting hot halo, which would otherwise have 
resulted from the IGM accretion;
the observed X-ray emission and absorption in the immediate vicinities
of the Galactic disk/bulge (\S~2.2) apparently arise from metal-rich
gas, or the product of the ongoing stellar feedback. The low-metallicity 
(hence radiation in-efficient) gaseous halo further alleviates 
the theoretical over-cooling problem of the IGM accretion. 
A detailed modeling of the scenario will be presented elsewhere.

\section{Acknowledgments}

I appreciate various useful comments from Z. Li, D. McCammon, and R. 
Shelton and thank my students and collaborators for their contributions to
the work as reviewed above, which is partly supported by NASA/CXC
under grants NNX06AB99G, NNG05GC69G, and GO5-6078X.


\end{document}